\documentclass[12pt]{article}

\usepackage{mathptmx}      

\usepackage{amsmath}
\usepackage{amssymb}
\usepackage{graphicx}
\usepackage{natbib}
\usepackage{color}
\usepackage{verbatim}
\usepackage{url}

\newcommand{\A}{A}
\newcommand{\B}{B}
\newcommand{\C}{C}
\newcommand{\D}{D}

\newcommand{\e}{e}
\newcommand{\s}{e}

\newcommand{\g}{\mu}    
\newcommand{\F}{\mathcal{G}}  
\newcommand{\G}{G}            
\newcommand{\var}{\sigma}
\newcommand{\pd}{\partial}
\newcommand{\R}{L}
\newcommand{\PP}{P}

\newcommand{\p}{p}
\newcommand{\Se}{S}    

\newcommand{\FRLM}{FRLM}
\newcommand{\FFCM}{FFCM}

\begin{document}
\title{Distributions of positive signals in pyrosequencing}


\author{Yong Kong\\
Department of Molecular
Biophysics and Biochemistry\\
W.M. Keck Foundation Biotechnology Resource Laboratory \\
Yale University\\
333 Cedar Street, New Haven, CT 06510\\
email: \texttt{yong.kong@yale.edu} }

%


\date{}

\maketitle


\begin{abstract}
Pyrosequencing is one of the 
important next-generation sequencing technologies.
We derive the distribution of the number of positive signals
in pyrograms of this sequencing technology 
as a function of flow cycle numbers
and nucleotide probabilities
of the target sequences.
As for the distribution of sequence length, we also
derive the distribution of positive signals for the fixed flow cycle model.
Explicit formulas are derived for the mean and variance
of the distributions.
A simple result for the mean of the distribution is that
the mean number of positive signals in a pyrogram
is approximately twice the number of flow cycles,
regardless of nucleotide probabilities.
The statistical distributions will be useful 
for instrument and software development for pyrosequencing
and other related platforms.
\end{abstract}

\section{Introduction} \label{S:intro}

The next-generation sequencing is transforming 
biological research in many aspects. Pyrosequencing is one of 
the important new sequencing platforms.
Compared with other next-generation sequencing technologies,
currently  pyrosequencing has the advantage of 
longer sequence read length,
which makes it possible for \emph{de novo} sequencing of new genomes
and other applications.

Pyrosequencing technology is based on  detection of  pyrophosphate (PPi)
that is released during DNA synthesis.
The four kinds of nucleotides (dATP, dCTP, dGTP, and dTTP) are added
iteratively in a pre-defined order, 
with one kind of nucleotide at a time.
For each nucleotide flow, 
if the nucleotide is complementary to the
DNA template, the nucleotide is incorporated
by polymerase.
A detectable light is then emitted 
with an intensity which is in theory proportional to the
number of incorporated nucleotides. 
If the added nucleotide is not complementary to the template,
ideally no light signal is emitted.
Because we know which nucleotide is added at each nucleotide flow, 
the DNA sequence of the
template can be determined by the presence or absence of the emitted light
and its intensity.
The detected light signals are usually presented as a \emph{pyrogram}
(also called \emph{flowgram}), 
in which the x-axis is the pre-determined nucleotide flows
and y-axis is the intensity of the emitted light at each nucleotide flow. 
Thus the pyrogram consists of both \emph{positive} and 
\emph{zero} signals: positive signals correspond to the emitted light signals,
which in turn reflect nucleotide incorporation activities,
while zero signals correspond to background noise with no
nucleotide incorporation.
In a pyrogram 
the positive signals are interspersed with zero signals.

To avoid unnecessary specification of the
detailed names of the four kinds of nucleotides, in the following we will
use $a$, $b$, $c$, and $d$ to represent any permutations of 
the usual nucleotides $A$, $C$, $G$, and $T$.
In Table~\ref{T:defs} the relations between nucleotide flow, 
flow cycle number $f$,
and the pyrogram of a sample sequence are shown.
The sample sequence shown in Table~\ref{T:defs} is
\begin{equation}  \label{E:seq}
  bddabaaad.
\end{equation}
If flow order is \{$a$-$b$-$c$-$d$\},
the pyrogram of above sample sequence is
\{0-1-0-2\}-\{1-1-0-0\}-\{3-0-0-1\}, or in short-hand notion
\begin{equation}  \label{E:pyro}
 0  1  0  2  1  1  0  0  3  0   0   1 ,
\end{equation}
where non-zeros are the positive signals.
In the first flow cycle ($f=1$),
the first three nucleotides $bdd$ are synthesized to give out
signals \{0-1-0-2\},
in the second flow cycle ($f=2$),
the next two nucleotides $ab$ are synthesized to give out
signals \{1-1-0-0\},
and in the third flow cycle ($f=3$),
the remaining nucleotides $aaad$ are synthesized to give out
signals \{3-0-0-1\}.
If we represent the positive signals using the upper case letters
of the corresponding nucleotide flow and ignore the zeros, 
we end up with the last row
in Table~\ref{T:defs}:
\begin{equation}  \label{E:runs}
  BDABAD.
\end{equation}
Sequence~\eqref{E:runs} represents the \emph{runs}
of the original Sequence~\eqref{E:seq}, which we will call as
\emph{r-seq} of the original sequence.
Here a \emph{run} in a random sequence is defined
as a stretch of consecutive identical element of one object
separated by elements of other objects or the end of the 
sequence.
The study of distributions of various runs
has a long history and has found important applications in
various fields~\citep{Balakrishnan2002,Kong2006,Kong2013}.
It is obvious that for any given sequence such as Sequence~\eqref{E:seq},
there is a unique corresponding r-seq, which is obtained 
by shrinking a stretch of consecutive identical letters (defined as
a \emph{run}) into a single letter.
The reverse is however not true. 
In fact, there are infinite number of sequences that can have 
Sequence~\eqref{E:runs} as their r-seqs. 
The first $B$ in Sequence~\eqref{E:runs}, for example,
can come from $b$, $bb$, $bbb$, etc.
The length of r-seq is clearly 
dependent on the context of the original sequence.
For example, if we just shuffle Sequence~\eqref{E:seq} into another sequence
\[
 aaaabbddd,
\]
the corresponding r-seq will be $ABD$.

Previously, we have studied the distribution of the sequence
length (of the original sequences) 
for pyrosequencing as well as the more general
case where nucleotide incorporation is probabilistic
~\citep{Kong2009,Kong2009b,Kong2012}.
In this paper, we are interested in the distribution of the length
of r-seqs:
what is the distribution of the number of the positive signals
in a pyrogram, for a given number of flow cycles?

\begin{table}
  \centering
\caption{
The relation between nucleotide flow, flow cycle number $f$, the pyrogram of
a sample sequence $bddabaaad$, and its corresponding positive signals.
}
\label{T:defs}
\begin{tabular}{ l|cccc|cccc |cccc }
\hline \hline
Flow cycle & \multicolumn{4}{|c|}{$f=1$} 
           & \multicolumn{4}{|c}{$f=2$} 
           & \multicolumn{4}{|c}{$f=3$} 
\\
Nucleotide flow 
    & $a$ & $b$ & $c$ & $d$ &
      $a$ & $b$ & $c$ & $d$ &
      $a$ & $b$ & $c$ & $d$ \\
\hline
Pyrogram         
  & 0 & 1 & 0 & 2 & 1 & 1 & 0 & 0 & 3 & 0  & 0  & 1 \\
Positive signals
  &   & $B$ &   & $D$ & $A$ & $B$ &   &   & $A$ &    &    & $D$ \\
\hline
\end{tabular}
\end{table}

The distribution of the positive signals in a pyrogram (the length
of r-seq),
besides being an interesting self-contained mathematical problem,
has important practical applications for algorithm and software development
of the pyrosequencing technology.
It has immediate implications for base-calling algorithms.
Currently,
base-callers for pyrosequencing use fixed pre-determined
thresholds to call the presence or absence of any bases.
Due to signal intensity variations and background noises,
base-calling errors (insertions and deletions) 
are inevitable by using fixed thresholds.
In fact, the higher amount of
insertion and deletion errors made during base-calling step
is a major drawback of pyrosequencing and similar sequencing technologies.
The availability of the distribution of the positive signals in the pyrograms
points to a potentially new direction for base-caller
algorithm development for pyrosequencing.
For example, one can envision that in a bootstrapping fashion,
the thresholds used to call the bases 
can be iteratively adjusted based on the 
deviation of the number of positive signals obtained by these thresholds
from the expected distribution, 
so that an optimal set of thresholds can be obtained that 
minimizes the difference. 
Another use of the theoretical distribution of positive signals
is to test base independence.
The distribution obtained in this paper
is based on the assumption that
the nucleotides in the target
sequence are independent of each other.
If the bases in the sequence are correlated, 
this assumption will be violated and the actual distribution
of the sequence will not obey the theoretical distribution.

The development of this paper 
follows similar route we took to obtain the distribution
of sequence length.  First we study the situation where the
number of positive signals is pre-defined as $r$.
We term this as \emph{fixed r-seq length model} (\FRLM).
This is analogous to the \emph{fixed sequence length model}
when the sequence length distributions were studied.
Within this model we first develop proper recursive relations,
then use generating functions (GFs) to solve these
recursive relations.
The GFs obtained then readily yield the mean and variance
for the distribution.
From \FRLM{} then we transform to the more realistic 
\emph{fixed flow cycle model}
(\FFCM), in which instead of fixing length of r-seq,
we fix the number of flow cycles $f$
and treat $r$ as a random variable.
This model is of more practical use. 
A simple and somewhat surprising result
coming from the distribution of $r$ is that the mean
of $r$ is approximately twice the number of flow cycles:
$\bar{r} \approx 2 f$ 
(Eq.~\eqref{E:avg_fixed_f} for \FRLM{} and
Eq.~\eqref{E:avg_ffcm} for \FFCM), regardless of  nucleotide probabilities
of the target sequences.
In other words, on average in a pyrogram about half of the nucleotide flows
has positive signals, and the other half has zero signal.  

The GF approach not only gives the exact distributions and 
various moments of the distributions (with mean and variance as special cases),
it also leads naturally to the limiting 
distributions \citep{Flajolet2009}.
All the distributions obtained previously from this series of work
as well as those derived in this paper
are asymptotically Gaussian, with the mean and variance
linear with the size of the system.

The paper is organized as follows.
First in the remaining of this \emph{Introduction} section
we  define the necessary notation.
Then we proceed to fixed r-seq length model (\FRLM).
There the recurrences and GF solution are presented.
Explicit formulas are derived for the mean and variance of the distributions,
for both fixed number of cycles and fixed number of positive signals.
Based on results from \FRLM{} we then obtain the results for \FFCM.
The analytical results of \FFCM{} are carefully compared with
the simulation results, which are posted.
In the last section we summarize and discuss the
results obtained so far for the pairwise relations
between $f$ (flow cycle), $r$ (r-seq length), 
and $n$ (sequence length), and point out the
non-transitive nature for the variances of these distributions.
In this paper, we will only focus on the case of
pyrosequencing, where the nucleotides are completely incorporated
in each nucleotide flow.
The case of incomplete nucleotide incorporation will be discussed
elsewhere.

\subsection{Notation and definitions}

As stated earlier,
we will
use $a$, $b$, $c$, and $d$ to represent any permutations of 
the usual nucleotides $A$, $C$, $G$, and $T$.
As shown in Sequence~\eqref{E:runs}, the upper case letters
$A$, $B$, $C$, and $D$ are used for the runs of nucleotides
of the corresponding lower case letters.
Throughout the paper we assume that the nucleotides in the target
sequence are independent of each other. 
The probabilities for the four nucleotides in the target sequence
are denoted as 
$p_a$, $p_b$, $p_c$, and $p_d$. 
In Table~\ref{T:defs} we define the \emph{flow cycle number} $f$
(the first row) to distinguish it from nucleotide flow cycle
(the second row).
A flow \emph{cycle}  
is the ``quad cycle'' of successive four nucleotides \{$abcd$\}.
The flow cycle number is denoted as $f$ in the following.
We assume that the flow cycle $f$ is always a complete cycle (i.e.,  
the number of nucleotide cycles is always a multiple of $4$).
We will use $r$ for the number of positive signals in a pyrogram.
It is identical to the total number of runs in the original sequence,
or equivalently,
the length of the corresponding r-seq~\citep{Kong2006,Kong2013}.
For example, the length of Sequence~\eqref{E:runs}
is the total number of runs in the original Sequence~\eqref{E:seq},
and is also the number of positive signals in the pyrogram~\eqref{E:pyro}
of the same sequence.

In the following we will frequently use the definition of
\emph{elementary symmetric functions} to express the results in 
more compact forms.
For our purpose the elementary symmetric functions with four variables 
are defined in
Eq.~\eqref{E:esf}, in terms of the nucleotide probabilities:
\begin{align} \label{E:esf}
\begin{aligned}
  \s_1 &= p_a + p_b + p_c + p_d ,                                    \\
  \s_2 &= p_a p_b + p_a p_c +  p_a p_d + p_b p_c + p_b p_d + p_c p_d , \\ 
  \s_3 &= p_a p_b p_c +  p_a p_b p_d + p_a p_c p_d + p_b p_c p_d ,     \\
  \s_4 &= p_a p_b p_c p_d .                                           
\end{aligned}
\end{align}
It's obvious that we have the constraint on
$\s_1$ as 
$\s_1 = p_a + p_b + p_c + p_d = 1$.
We also need the following definition, 
which represents a value that is proportional to the probability of
\emph{one} positive signal
of a particular nucleotide, with the number of
nucleotides in the signal ranging from $1$ to infinity:
\begin{equation} \label{E:mu}
  \g_i = \sum_{j=1}^\infty p_i^j = \frac{p_i}{1 - p_i} ,
 \qquad i = a,b,c,d .
\end{equation}

We'll frequently extract coefficients from the expansion of GFs.
If $h(x)$ is a series in powers of $x$, then we use the notation
$[x^n]h(x)$ to denote the coefficient of $x^n$ in the series.
Similarly, we use $[x^n y^m]h(x,y)$ to denote the 
coefficient of $x^n y^m$ in the bivariate $h(x, y)$.

\section{Fixed r-seq length model (\FRLM)} 
\label{S:FRLM}
In this model we assume that the number of positive signals $r$ is fixed.  
Let $L_i(f, r)$ ($i=a$, $b$, $c$, and $d$) denote 
the probability (up to a normalization
factor, see below) 
that a r-seq of length $r$ and ending with nucleotide $i$
is synthesized in the first $f$ flow cycles, 
with the $r$-th positive signals synthesized in flow cycle $f$.
Symbolically we can enumerate the first few terms for small $r$
as in Table~\ref{T:fixed_r}.

\begin{table}
\caption{The first few terms for \FRLM.
Here $\mathcal{S}_2=AB+AC+BC+AD+BC+CD$.
Note that non-zero terms of $r=3$ and $f=3$ are not shown:
$CBA+DBA+DCA$ for nucleotide flow $a$, and $DCB$ for nucleotide flow $b$.
}
\label{T:fixed_r}
\scriptsize
\begin{tabular}{ c|cccc|cccc }
\hline \hline
 & \multicolumn{4}{|c|}{$f=1$} 
 & \multicolumn{4}{|c}{$f=2$} \\
$r$ & $a$ & $b$ & $c$ & $d$ &
   $a$ & $b$ & $c$ & $d$ \\
\hline
$1$ & $\A$ & $\B$    & $\C$        & $\D$ & $0$ & $0$ & $0$ & $0$  \\
$2$ & 0    & $\A \B$ & $(\A+\B)\C$ & $(\A+\B+\C)\D$ & 
        $(\B+\C+\D)\A$ & $(\C+\D) \B$   & $\D \C$   & 0 \\
$3$ & 0    &  0      & $\A \B \C$    & $(\A \B + \A \C + \B \C) \D $
    & $\mathcal{S}_2 \A $ 
    & $(\mathcal{S}_2 + \C\A + \D\A) \B$
    & $(\mathcal{S}_2 + \A\D + \B\D) \C$
    & $ \mathcal{S}_2 \D$
\\
\hline
\end{tabular}
\end{table}

\subsection{Recurrences}
For the sequencing model specified
(complete nucleotide incorporation for each nucleotide flow),
the following recurrence relations can be established:
\begin{align} \label{E:rec}
\begin{aligned}
 L_a(f+1, r+1) &= \left[               L_b(f, r)   + L_c(f, r  ) + L_d(f, r)
                  \right] \g_a ,       \\
 L_b(f+1, r+1) &= \left[ L_a(f+1, r)               + L_c(f, r)   + L_d(f, r)
                  \right] \g_b ,       \\
 L_c(f+1, r+1) &= \left[ L_a(f+1, r) + L_b(f+1, r)               + L_d(f, r)
                  \right] \g_c ,       \\
 L_d(f, r+1)   &= \left[ L_a(f, r)   + L_b(f, r)   + L_c(f, r)   
                  \right] \g_d ,
\end{aligned}
\end{align}
where $\g_i$ is defined in Eq.~\eqref{E:mu}.

The first of these recurrences is true because
for any sequence ending with nucleotide $a$ and the length of its r-seq as
$r+1$ to be synthesized in the first $f+1$ flow cycles
with the $(r+1)$-th run of nucleotides synthesized in flow cycle $f+1$,
the r-seq has to be of the form $\boxed{\cdots} XA$, where $X=B$, $C$, or $D$.
So $X$ has to be synthesized in one of the nucleotide flows
in flow cycle $f$.
Which  nucleotide flow $X$ is synthesized
depends on what $X$ is ($B$, $C$, or $D$).
The length of r-seq $\boxed{\cdots} X$ is obviously $r$.
The second recurrence is true because in this case the r-seq must be in the
form of  $\boxed{\cdots} XB$, with $X=A$, $C$, or $D$.  If $X=A$, 
this run of $a$'s must be synthesized in the nucleotide flow $a$ of
flow cycle $f+1$;
if $X=C$ or $D$ 
this run of $c$'s ($d$'s) must be synthesized in the nucleotide flow $c$ ($d$)
 of the previous flow cycle, the flow cycle $f$.
The other two recurrences are true for similar reasons.

The recurrences of Eq.~\eqref{E:rec} cannot be solved in closed forms.
However, their GFs can be solved in compact forms.

\subsection{Generating functions}
The GFs of $L_i(f, r)$ are defined as
\[
 G_i (x, y) 
 = \sum_{r=1}^\infty \sum_{f=1}^\infty L_i(f, r) x^f y^r,
 \qquad i = a, b, c, d.  
\]
By using proper initial conditions,
these GFs are solved as 
\begin{align} \label{E:Gi}
\begin{aligned}
  G_a &= \frac{\g_a x y (1 + \g_b xy) (1 + \g_c xy) (1 + \g_d xy)}{H} , \\
  G_b &= \frac{\g_b x y (1 + \g_a  y) (1 + \g_c xy) (1 + \g_d xy)}{H} , \\
  G_c &= \frac{\g_c x y (1 + \g_a  y) (1 + \g_b  y) (1 + \g_d xy)}{H} , \\
  G_d &= \frac{\g_d x y (1 + \g_a  y) (1 + \g_b  y) (1 + \g_c  y)}{H} 
\end{aligned}
\end{align}
where
\begin{equation} \label{E:H}
 H = 1 - \Se_2 x y^2 - \Se_3 x(1+x) y^3 - \Se_4 x(1+x+x^2) y^4 ,
\end{equation}
and $\Se_i$ are the elementary symmetric function of $\g_i$:
 \begin{align*}
   \Se_1 &= \g_a + \g_b + \g_c + \g_d          , \\
   \Se_2 &= \g_a \g_b + \g_a \g_c + \g_a \g_d 
   + \g_b \g_c + \g_b \g_d + \g_c \g_d         , \\
   \Se_3 &= \g_a \g_b \g_c + \g_a \g_b \g_d + \g_a \g_c \g_d + \g_b \g_c \g_d 
                                              , \\
   \Se_4 &= \g_a \g_b \g_c \g_d .
 \end{align*}

The value of $L_i(f, r)$ can be obtained from their corresponding GF $G_i(x, y)$
by extracting the appropriate coefficients.
By using the notation we introduced earlier
we have $L_i(f, r) = [x^fy^r] G_i (x, y)$, $i=a$, $b$, $c$, and $d$.

From the expressions of the GFs we can see that they are not symmetric
with respect to $\g_i$, $i=a$, $b$, $c$, and $d$.
If we only consider the nucleotide flows  
that end up in the same ``quad cycle'' (see Table~\ref{T:defs}),
then we can add the four GFs together
to obtain
\begin{align} \label{E:G_sum}
G (x, y) &= G_a + G_b + G_c + G_d \notag \\
  &= \frac{xy \left[ \Se_1 + \Se_2 (1+x)y + \Se_3 (1+x+x^2) y^2 
      + \Se_4 (1+x+x^2+x^3) y^3
     \right]}{H} .
\end{align}
The expression of $G(x, y)$ is symmetric
with respect to $\g_i$, and hence to the nucleotide probabilities $p_i$, 
since all the 
parameters involved are encapsulated in $\Se_i$, $i=1, 2,3,4$, 
the elementary symmetric functions of $\g_i$.
In the following we focus on this symmetric GF $G (x, y)$.


\subsection{Normalization factors} \label{SS:norm}

The values of $L_i(f, r)$ from Eq.~\eqref{E:Gi}
or $L(f, r)$ from Eq.~\eqref{E:G_sum} do not added up to $1$ 
when either $r$ or $f$ is fixed,
so for $L(f, r)$ to become probability they have to be normalized.
The need for normalization can also be seen from Table~\ref{T:fixed_r}.
The probabilities of the four entries in the first row are given by
$\g_i/\sum_{j \in \{ a,b,c,d \}} \g_j$, for $i=a, b, c, d$.
These four probabilities sum up to $1$.
For the second row and rows below, however,
the sum of the probabilities of each row does not equal to $1$.
For the second row, only when entries of $A^2$, $B^2$, $C^2$, and $D^2$
are added can the sum of probabilities become $1$.
By definition, the r-seq length of these entries is $1$ instead of $2$
(from r-seq point of view, they are equivalent to $A$, $B$, $C$, and $D$),
so they should not be included in the second row.

By setting  $x=1$ in $G(x, y)$ of Eq.~\eqref{E:G_sum}, 
we get $G(1, y) = \sum_{r=1}^\infty [ \sum_{f=1}^\infty L(f, r) ]  y^r = 
\sum_{r=1}^\infty u(r) y^r$.
The normalization factor $u(r)$ at a fixed number of positive
signals $r$ can be obtained as
\[
 u(r) =  \sum_{f=1}^\infty L(f, r) = [y^r] G(1, y).
\]
Similarly, 
by setting  $y=1$ in $G(x, y)$ of Eq.~\eqref{E:G_sum}, 
we get 
the normalization factor $v(f)$ at fixed cycle number $f$ as
\[
 v(f) = \sum_{r=1}^\infty L(f, r) = [x^f] G(x, 1).
\]

It can be shown that both $x=1$ and $y=1$ are roots of $H$ in Eq.~\eqref{E:H}.
For $u(r)$, $y=1$ is the dominant part in the expansion of $G(1, y)$, 
and  this dominant term in the expansion leads to
\begin{equation} \label{E:u}
 u(r) \approx \frac{1}{2 e_2} .
\end{equation}
When the sequences have equal nucleotide probabilities 
($p_a = p_b = p_c = p_d = 1/4$), Eq.~\eqref{E:u} is exact.
When the nucleotide probabilities are unequal, 
there are  extra terms in the expression, 
but they are so small
for even moderate $r$ that 
practically they can be ignored.
The bigger $r$, the smaller contributions of these
extra terms.
For clarity reason, these small terms are not shown here.
The following example shows how $u(r)$ in 
Eq.~\eqref{E:u} approaches the true value as $r$ increases.
When 
$p_a = 1/3$,
$p_b=1/11$, 
$p_c=100/231$, 
and 
$p_d=1/7$, we have the following expansion of $G(1, y)$,
\begin{multline*}
 G(1, y) = 
  1.530025445 y
+ 1.470483461 y^2 + 1.493450006 y^3 + 1.482494027 y^4 \\
+ 1.488505721 y^5
+ 1.484956420 y^6 + 1.487123159 y^7 + 1.485781400 y^8 \\
+ 1.486617232 y^9
+1.486095292 y^{10}  + \cdots.
\end{multline*}
The value $1/2 e_2 = 1.486296028$ is converged upon to the ninth decimal place
when $r \ge 38$.

For $v(f)$, $x=1$ is the dominant part in the expansion of $G(x, 1)$,
and $v(f)$ is approximately given by
\begin{equation} \label{E:v}
 v(f) \approx \frac{1}{e_2} .
\end{equation}
Eq.~\eqref{E:v} approaches the true value quickly as $f$ increases.
For example,
for 
$p_a = 1/3$,
$p_b=1/11$, 
$p_c=100/231$, 
and 
$p_d=1/7$, we have the following expansion of $G(x, 1)$:
\begin{multline*}
 G(x, 1) = 
   2.394465649 x   + 2.934312576 x^2 + 2.975097540 x^3 + 2.972495921 x^4 \\
 + 2.972589150 x^5 + 2.972593062 x^6 + 2.972591938 x^7 + 2.972592066 x^8 \\
 + 2.972592056 x^9 + 2.972592056 x^{10} + \cdots,
\end{multline*}
which shows that $v(f)$ converges to $1/e_2$ at ninth decimal place
when $f \ge 9$.

\subsection{Mean and variance}
The availability of GFs makes it easy to derive various moments of the
distribution, including the mean and variance.

When the number of positive signals $r$ is fixed,
the mean and variance of the number of flow cycles $f$ are given by
\begin{align}
 \bar{f} (r) &= \frac{1}{u(r)} 
     [y^r]\frac{\pd G(x, y) } {\pd x} \Big |_{x=1}   , 
 \label{E:avg_fixed_r}\\
 \var^2_f (r)  &= \frac{1}{u(r)} 
     [y^r]\frac{\pd^2 G(x, y) } {\pd x^2} \Big |_{x=1} 
 + \bar{f}(r) - \bar{f}^2(r) . \label{E:var_fixed_r}
\end{align}

When the number of flow cycles is fixed,
the mean and variance of the number of positive signals $r$ are given by
\begin{align}
 \bar{r} (f) &= \frac{1}{v(f)} 
     [x^f]\frac{\pd G(x, y) } {\pd y} \Big |_{y=1}  ,  
     \label{E:avg_fixed_cycle}\\
 \var^2_r (f)  &= \frac{1}{v(f)} 
     [x^f]\frac{\pd^2 G(x, y)} {\pd y^2} \Big |_{y=1} 
 + \bar{r}(f) - \bar{r}^2(f) .  \label{E:var_fixed_cycle}
\end{align}

\subsection{Distribution of the number of positive signals
as a function of flow cycle number} 

At a particular number of flow cycles $f$,
the distribution of the number of positive signals
can be calculated from Eq.~\eqref{E:G_sum} as
\begin{equation} \label{E:prf}
 \frac{[x^f y^r] G(x, y)}{[x^f] G(x, 1)} 
 = \frac{1}{v(f)} [x^f y^r] G(x, y) .
\end{equation}
From Eqs.~\eqref{E:avg_fixed_cycle} and \eqref{E:var_fixed_cycle}
the mean and variance of $r$ at the fixed $f$
 can be calculated as
\begin{subequations} \label{E:avg_var_fixed_f}
 \begin{align}
 \bar{r}(f)   & \approx 2 f - \frac{e_2 - e_3}{e_2} , \label{E:avg_fixed_f} \\
 \var^2_r (f) & \approx \frac{2 e_3}{e_2} f + 
 \frac{2 e_2^3 + 5 e_3^2 - 3 e_2 e_3 - 4 e_2 e_4}{e_2^2} . \label{E:var_fixed_f} 
 \end{align}
\end{subequations}
In Eq.~\eqref{E:avg_var_fixed_f} we only keep the dominant part of
$x=1$ in the series expansion of 
Eqs.~\eqref{E:avg_fixed_cycle} and \eqref{E:var_fixed_cycle}.
The ignored non-dominant terms go to zero as $f$ increases.
For the special case when $p_a = p_b = p_c = p_d = 1/4$, we have
 \begin{align*}
   \bar{r}(f)   & \approx 2 f - \frac{5}{6} , \\
   \var^2_r (f) & \approx \frac{1}{3} f + \frac{25}{72} .
 \end{align*}

It is interesting to note that, as shown in Eq.~\eqref{E:avg_fixed_f}, 
on average
the number of positive signals is about twice that of
the flow cycles, regardless of
the nucleotide probabilities of the target sequence.
 In other words, the numbers of positive and zero
signals in a pyrogram are about equal to each other on average.
From Eq.~\eqref{E:avg_var_fixed_f}
we see that both of the mean and variance of $r$
increase linearly with the flow cycle number $f$.
This linear growth of both mean and variance with the size of the system
also happens in the distributions of the sequence length~\citep{Kong2009,Kong2009b,Kong2012}.
In fact, a large number combinatorial systems governed by meromorphic GFs
have this property~\citep[Chapter IX]{Flajolet2009}.

\begin{figure}
  \centering
  \includegraphics[angle=270,width=\columnwidth]{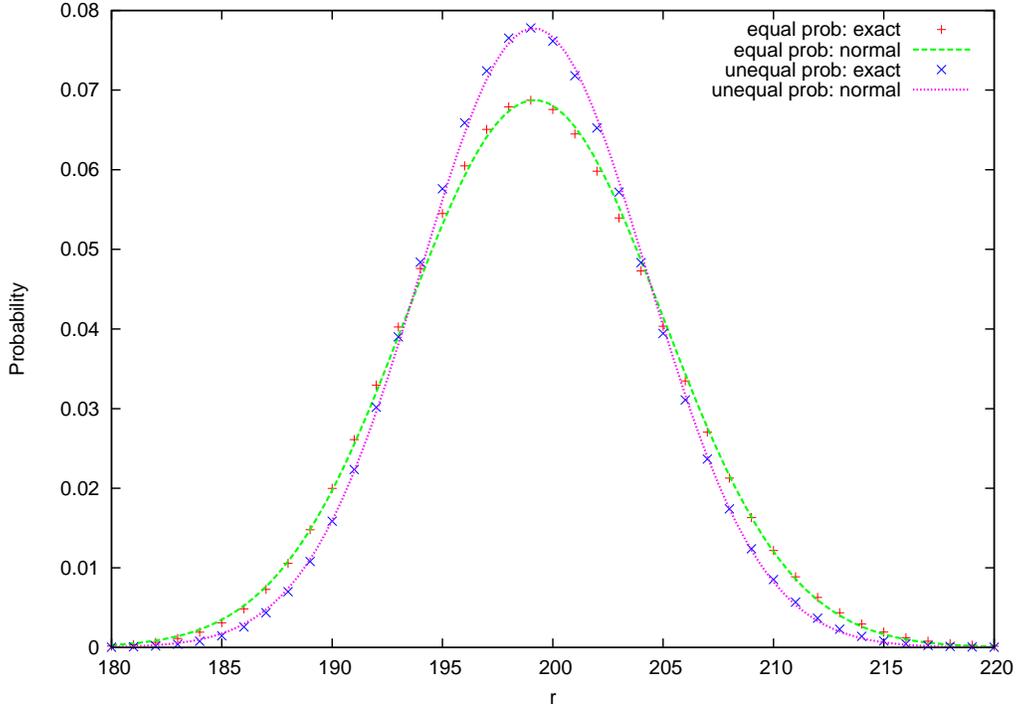}
  \caption{
    Distributions of positive signals when 
    flow cycles $f=100$, for both equal 
    and unequal nucleotide probabilities.
    The unequal nucleotide probabilities used here are
    $p_a=1/3=0.333$, $p_b=1/11=0.091$, $p_c = 100/231=0.433$,
    and $p_d=1/7=0.143$.
    The exact distributions are calculated from Eqs.~\eqref{E:prf} and
    \eqref{E:G_sum}.
    The continuous curves are 
    the normal distributions $N({\bar r} (f), \var_r^2 (f))$
    of the same mean and variance as those of the exact distributions,
    where ${\bar r} (f)$ and $\var_r^2 (f)$ are calculated
    from Eqs.~\eqref{E:avg_fixed_f} and \eqref{E:var_fixed_f}.
    The two normal distributions shown here are
    $N(199.167, 33.681)$ and $N(199.130, 26.316)$, 
    for equal and unequal nucleotide probabilities,
    respectively.
    \label{F:fixed_f}} 
\end{figure}

The GF (Eq.\eqref{E:G_sum}) leads naturally to  asymptotic distributions
according to the singularity perturbation theory detailed in 
~\citet[Chapter IX]{Flajolet2009}.  The theory claims that 
the limiting distribution is Gaussian, with mean and variance
linear with the size of the system.
In Figure~\ref{F:fixed_f}
the distributions of  positive signals when $f=100$ are plotted,  
for both equal 
and unequal nucleotide probabilities.
The unequal nucleotide probabilities used here are
arbitrarily chosen as
$p_a=1/3=0.333$, $p_b=1/11=0.091$, $p_c = 100/231=0.433$,
and $p_d=1/7=0.143$.
The exact distributions are calculated from Eqs.~\eqref{E:prf} and
\eqref{E:G_sum}.
Also shown in  Figure~\ref{F:fixed_f}
in continuous curves are 
the normal distributions $N({\bar r} (f), \var_r^2 (f))$
of the same mean and variance as those of the exact distributions,
where ${\bar r} (f)$ and $\var_r^2 (f)$ are calculated
from Eqs.~\eqref{E:avg_fixed_f} and \eqref{E:var_fixed_f}.
The two normal distributions shown here are
$N(199.167, 33.681)$ and $N(199.130, 26.316)$, 
for equal and unequal nucleotide probabilities,
respectively.
It is clear that the exact distributions can be approximated quite accurately
by the
normal distributions with the same mean and variance, for both
equal and unequal nucleotide probabilities.

By using the constraint $\s_1 = p_a + p_b + p_c + p_d = 1$,
it can be shown that when the four nucleotides
have equal probability $1/4$,
the variance of the distribution reaches its maximum. 
In other words, for a given flow cycle $f$, the distribution
of the number of positive signals for 
sequences with equal nucleotide probability
is broader than  
the distribution from sequences with unequal nucleotide probabilities.
In fact, the coefficient of $f$ in $\var^2_r (f)$, $2 e_3 / e_2$ in
Eq.~\eqref{E:var_fixed_f}, reaches its maximum of $1/3$
when the four nucleotides
have equal probability $1/4$.

\subsection{Distribution of the number of flow cycles 
as a function of number of positive signals} 
To be complete, 
we derive here the distribution of the number of flow cycles $f$
at fixed number of positive signals $r$.
The exact probability is expressed as
\begin{equation} \label{E:pfr}
 \frac{[x^f y^r] G(x, y)}{[y^r] G(1, y)} .
\end{equation}
By using Eqs.~\eqref{E:avg_fixed_r} and \eqref{E:var_fixed_r}
the mean and variance of $f$ at the fixed $r$
 can be calculated as
\begin{subequations}
 \begin{align}
 \bar{f} (r)  &= \frac{1}{2} r + \frac{1}{2} , \label{E:f_r_avg} \\
 \var^2_f (r) & \approx  \frac{e_3}{4 e_2} r 
   + \frac{2 e_2^3 + 3 e_3^2 - 2 e_2 e_3 - 4 e_2 e_4}{4 e_2^2} 
 \label{E:f_r_var} .
 \end{align}
\end{subequations}
It should pointed out that Eq.~\eqref{E:f_r_avg}
is \emph{exact},
while for Eq.~\eqref{E:f_r_var}, only the dominant term corresponding to $y=1$
is kept while non-dominant terms are ignored.

It is interesting to note that like the mean of $r$,
the mean of $f$
does not depend on the nucleotide probabilities.
Again both the mean and variance vary linearly with the
the number of positive signals $r$,
as predicted by the singularity perturbation theory~\citep{Flajolet2009}.
When the nucleotide probabilities are equal ($p_i=1/4$),
the variance in Eq.~\eqref{E:f_r_var} becomes
\begin{equation}
 \var^2_f (r) \approx \frac{1}{24} r + \frac{11}{96} .
\end{equation}
In this equal nucleotide probability case
the exact expression of $\var^2_f (r)$ can be written as
\[
 \var^2_f (r) = \frac{1}{24} r + \frac{11}{96}
 +\frac{15}{32}  \left( -\frac{1}{3} \right)^r  ,
\]
with $15/32 \times (-1/3)^r$ approaching zero as $r$ becomes
large.

\begin{figure} 
  \centering
  \includegraphics[angle=270,width=\columnwidth]{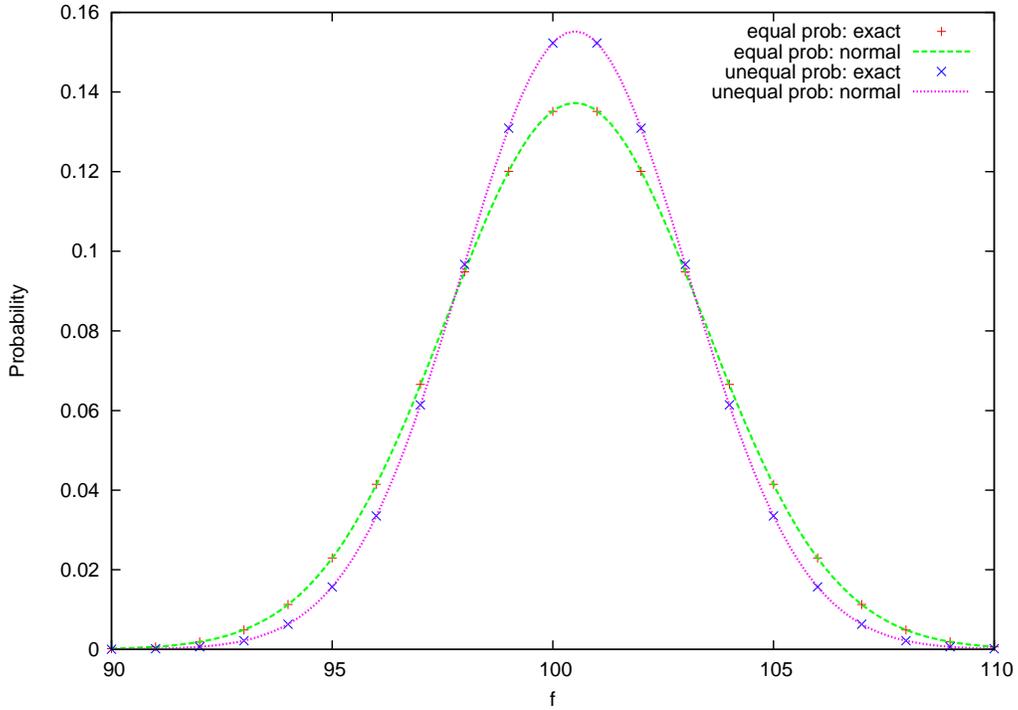}
  \caption{
    Distributions of the number of flow cycles 
    when the number of positive signals 
    $r=200$, for both equal 
    and unequal nucleotide probabilities.
    The unequal nucleotide probabilities used here are
    the same as in Figure~\ref{F:fixed_f}.
    The exact distributions are calculated from Eqs.~\eqref{E:pfr} and
    \eqref{E:G_sum}.
    The continuous curves are 
    the normal distributions $N({\bar f} (r), \var_f^2 (r))$
    of the same mean and variance as those of the exact distributions,
    where ${\bar f} (r)$ and $\var_f^2 (r)$ are calculated
    from Eqs.~\eqref{E:f_r_avg} and \eqref{E:f_r_var}.
    The two normal distributions shown here are
    $N(100.5, 8.448)$ and $N(100.5, 6.603)$, 
    for equal and unequal nucleotide probabilities,
    respectively.
    \label{F:fixed_r}} 
\end{figure}

Based on the singularity perturbation theory~\citep{Flajolet2009},
the limiting distribution is again Gaussian.
In Figure~\ref{F:fixed_r}
the distributions of flow cycles 
when 
$r=200$ are plotted for both equal 
and unequal nucleotide probabilities.
The unequal nucleotide probabilities used here are
the same as in Figure~\ref{F:fixed_f}.
The exact distributions are calculated from Eqs.~\eqref{E:pfr} and
\eqref{E:G_sum}.
Also shown in Figure~\ref{F:fixed_r}
in continuous curves are 
the normal distributions $N({\bar f} (r), \var_f^2 (r))$
of the same mean and variance as those of the exact distributions,
where ${\bar f} (r)$ and $\var_f^2 (r)$ are calculated
from Eqs.~\eqref{E:f_r_avg} and \eqref{E:f_r_var}.
The two normal distributions shown here are
$N(100.5, 8.448)$ and $N(100.5, 6.603)$, 
for equal and unequal nucleotide probabilities,
respectively.
As we can see from Figure~\ref{F:fixed_r},
the exact distributions can be approximated accurately by the
normal distributions with the same mean and variance, for both
equal and unequal nucleotide probabilities.

\section{Fixed flow cycle model (\FFCM)}

In this model, instead of fixing the r-seq  length $r$,
we fix the number of flow cycles $f$.
This is a more natural model than \FRLM.
As before~\citep{Kong2012},
we assume that the sequence is generated by a random process 
that can create sequences with infinite length, only the first part of which
will be sequenced by $f$ flow cycles.
Let $\PP_i(f, r)$  ($i=a, b, c, d$) denote the probability 
that $f$ flow cycles will 
synthesize a sequence whose r-seq is of length $r$ and
with the last incorporated nucleotide being $i$,
and $\PP(f, r)$ be the sum of $\PP_i(f, r)$:
$\PP(f, r) = \sum_{i \in \{a,b,c,d\}} \PP_i(f, r)$.
Let $\F_i(x, y)$ be the GF of $\PP_i(f, r)$:
\[
  \F_i(x, y) = \sum_{r=1}^\infty \sum_{f=1}^\infty \PP_i(f, r) x^f y^r
\]
and $\F(x, y)$ GF of 
$\PP(f, r)$: $\F(x, y) = \sum_{i \in \{a,b,c,d\}} \F_i(x, y)$.
Using the same argument as before~\citep{Kong2012},
we have the following relation between $\PP_i(f, r)$ and $\R_i(f, r)$:
\begin{align}
\begin{aligned}
\label{E:pyro_rel}
 \PP_a(f, r) &= 0 ,                              \\
 \PP_b(f, r) &= \R_b(f, r) p_a ,                 \\
 \PP_c(f, r) &= \R_c(f, r) (p_a + p_b)  , \      \\
 \PP_d(f, r) &= \R_d(f, r) (p_a + p_b + p_c). 
\end{aligned}
\end{align}
The first part of Eq.~\eqref{E:pyro_rel} is true because
the first $f$ flow cycles can never synthesize a r-seq
ending with $a$ and of length $r$.
Based on our sequence model with the assumptions
of complete nucleotide incorporation
and infinite sequence length,
if the next nucleotides to be synthesized are of type $a$,
then they will be kept synthesized continuously within the nucleotide flow
$a$ of flow cycle $f$ until a nucleotide that is not
of type $a$ (say, a nucleotide of type $b$) comes up.
Then this nucleotide will be synthesized within the nucleotide flow
$b$ (if it is of type $b$) of flow cycle $f$, making the length of
the r-seq $r+1$.
The second part of Eq.~\eqref{E:pyro_rel} is true because
if the first $f$ flow cycles are to synthesize a r-seq of
length $r$ and ending with $b$, then the first $r$ positive signals of the
sequence must be synthesized in the first $f$ flow cycles
(which happens with probability proportional to $\R_b(f, r)$),
and no more positive signal must be synthesized in the $f$-th
flow cycle.
Only when the next base to be synthesized is of type $a$ can 
the synthesis in flow cycle $f$ be stopped; otherwise the synthesis 
will continue at the nucleotide flow
$b$ if the next base is $b$, or the number of positive signals 
becomes $r+1$ if the next base is $c$ or $d$.
The last two parts of Eq.~\eqref{E:pyro_rel} are true for similar
reasons.

From Eq.~\eqref{E:pyro_rel} we can obtain similar relations
between GFs $\F_i(x, y)$
and $\G_i(x, y)$:
\begin{align}
\label{E:pyro_F}
\begin{aligned}
 \F_a(x, y) &= 0 ,                               \\
 \F_b(x, y) &= \G_b(x, y) p_a ,                   \\
 \F_c(x, y) &= \G_c(x, y) (p_a + p_b)  ,          \\
 \F_d(x, y) &= \G_d(x, y) (p_a + p_b + p_c) . 
\end{aligned}
\end{align}
The GF $\F(x, y)  = \sum_{i \in \{a,b,c,d\}}  \F_i(x, y)$ 
thus calculated using Eq.~\eqref{E:Gi} and above relations
gives the distribution of $r$ for \FFCM.
It's explicit form, however, is not as compact as Eq.~\eqref{E:G_sum}
for \FRLM{} so is not shown here.
On the other hand, $\F(x, y)$ shows a nice property that Eq.~\eqref{E:G_sum}
does not have:
\[
 \F(x, 1) = \frac{x}{1-x} = x + x^2 + x^3 + \cdots,
\]
which shows that $\PP(f, r)$ generated by $\F(x, y)$ is a 
probability distribution without the need for normalization as
in \FRLM{} in the previous sections: for any $f$,
$\sum_{r=1}^\infty \PP(f, r) = 1$. 

From $\F(x, y)$ the mean and variance can be calculated as
\begin{subequations} \label{E:avg_var_ffcm}
 \begin{align}
 \bar{r} (f)    & \approx 2 f + \frac{ e_3 - e_2  + \p_a e_2 + w } {e_2} ,
 \label{E:avg_ffcm} \\
 \sigma^2_r (f) &  \approx
 \frac{2 e_3}{e_2} f +
 \frac{3 e_3^2 - e_2 e_3 - 2 e_2 e_4   
   + \p_a (1-\p_a) e_2^2 -  w^2 + (2 \p_b \p_d \p_c^2 + w) e_2}{e_2^2}
 \label{E:var_ffcm}
 \end{align}
\end{subequations}
where
\[
 w = \p_b \p_c \p_d + \p_c \p_b^2 + \p_d \p_b^2 + \p_d \p_c^2 .
\]
We note that the mean and variance for \FFCM{} (Eq.~\eqref{E:avg_var_ffcm}) 
have the same
coefficients of $f$ as in \FRLM{} (Eq.~\eqref{E:avg_var_fixed_f}).  
They only differ in the constant terms.
For \FFCM{}, the constant terms are no longer symmetric for 
nucleotide probability $p_i$.
For the special case of equal nucleotide probabilities, we have
\begin{align*}
 \bar{r} (f)    & \approx 2 f - \frac{5}{12}, \\
 \sigma^2_r (f) & \approx \frac{1}{3} f + \frac{35}{144} .
\end{align*}

Since $\F_i(x, y)$'s are linear transform of $\G_i(x, y)$'s, 
$\F(x, y)$ is still a bivariate GF that obeys 
the singularity perturbation theory~\citep{Flajolet2009}.
The limiting distribution of $\PP(f,r)$ is Gaussian and
its mean and variance are linear function of $f$, as
Eq.~\eqref{E:avg_var_ffcm} shows.

\section{Simulation}
To check the analytical results of \FFCM{} developed in this paper,
we developed a simulation program written in C programming language.
The program can be found in
\url{http://graphics.med.yale.edu/sbs/}.
All the results of this paper have been carefully checked
against simulation results.
Using seed $s=12345$ and repetition of $2 \times 10^7$,
for $f=100$,
    $p_a=0.333333333$,
    $p_b=0.090909091$,
    $p_c=0.432900433$,
    $p_d=0.142857143$,
the simulation gives the mean and variance as 
$199.574162$ and $26.209019$
respectively.
The theoretical results from Eq.~\eqref{E:avg_var_ffcm} gives
$199.573625$ and $26.215222$ for \FFCM, which are closer to the 
simulation results than \FRLM, which gives 
$199.129853$ and $26.315978$ as mean and variance
(Eq.~\eqref{E:avg_var_fixed_f}).
This simulation results are posted in the same website as
the simulation program itself.

\section{Some remarks on relations between $f$, $n$, and $r$}
In this paper 
we have derived the relation between $f$, the number of flow cycles,
and $r$, the number of positive signals in pyrograms.
Previously, we have obtained the relation between $f$ and $n$, 
the sequence length~\citep{Kong2009,Kong2012}.
The distribution of the number of runs $r$ in random sequences
with a fixed length $n$
 has also been derived
\citep{Kong2013}.  Here we summarize these relations below.
The mean and variance of $n$ when $f$ is fixed for \FFCM{}
~\citep{Kong2012}
\begin{align} \label{E:n_f_avg}
\begin{aligned}
  \bar{n} (f) & \approx  \frac{f}{\e_2} + \frac{\e_3}{\e_2^2} - 1, \\
  \var^2 (f)  & \approx \frac{\e_2 + 2\e_3 - 3 \e_2^2}{\e_2^3} f
 - \frac{\e_2^2 \e_3  -5 \e_3^2 + 4 \e_2 \e_4 }{\e_2^4}.  
\end{aligned}
\end{align}
The mean and variance of $r$ when $f$ is fixed for \FFCM{} derived in
this work are given by
Eq.~\eqref{E:avg_var_ffcm}.
The mean and variance of $r$ for a random sequence with a length of $n$
are given by ~\citep{Kong2013}
\begin{align}
 \bar{r} (n) &= (1 - P_2) n +  P_2                                  \notag  \\
             &= 2 \s_2 n + 1 - 2 \s_2,
 \label{E:r_n_avg} \\
 \var_r^2(n) &= (P_2 + 2 P_3 - 3 P_2^2) n  - (P_2 + 4 P_3 - 5 P_2^2) \notag \\
  &= (4 \s_2 + 6 \s_3 - 12 \s_2^2) n - (6 \s_2 + 12 \s_3 - 20 \s_2^2)
 \label{E:r_n_var},
\end{align}
where $P_j = \sum_{i \in \{a,b,c,d \}} p_i^j$.

The average of these three distributions are transitive, if we ignore
the small constant terms.
If we use Eq.~\eqref{E:n_f_avg} to calculate the average 
of sequence length $n$ as a function of $f$,
we have $\bar{n}(f) \approx f/\s_2$.
If we put it in Eq.~\eqref{E:r_n_avg} 
$\bar{r}(n) \approx 2 \s_2 n$, we obtain
$\bar{r} (f) \approx 2 f$, which is correct based on the results from this work
(Eq.~\eqref{E:avg_ffcm}.
For the variances, however, these transitive relations do not hold.
For example, for equal probabilities $p_i = 1/4$, the means and variances
for the three distributions are summarized below.
The first two equations are from \FFCM.
\begin{align}
  {\bar n} (f)  & \approx \frac{8}{3} f - \frac{5}{9}  &
  \var_n^2 (f)  & \approx \frac{40}{27} f + \frac{20}{81} ,   
\label{E:avg_n_f} \\
   \bar{r}(f)   & \approx 2 f - \frac{5}{12}           &
   \var^2_r (f) & \approx \frac{1}{3} f + \frac{35}{144} 
   \label{E:var_r_f} \\
\bar{r} (n) &= \frac{3}{4} n + \frac{1}{4} ,           &
  \var_r^2(n)   &= \frac{3}{16} n - \frac{3}{16} \label{E:var_r_n}.
\end{align}
If we use Eq.~\eqref{E:avg_n_f} to get the average of sequence length
${\bar n}(f) \approx 8f/3 $ and put it into Eq.~\eqref{E:var_r_n},
we end up with
$\var_r^2(f) \approx f/2 $, which is different from the correct
expression shown in
Eq.~\eqref{E:var_r_f} as $\var_r^2(f) \approx f/3 $.
In fact, the variance $\var_r^2(f)$ will always be overestimated
if we use Eq.~\eqref{E:n_f_avg}  to get the average of sequence length
$n$ as $\bar{n}(f) \approx f/\s_2$
and use it in Eq.~\eqref{E:r_n_var} to estimate $\var_r^2(f)$
as $\var_r^2(f) \approx (4 \s_2 + 6 \s_3 - 12 \s_2^2) f/ \s_2$.
By using the constrain $\s_1 = 1$, it can be shown that $(4 \s_2 + 6 \s_3 - 12 \s_2^2)/ \s_2$ is always more than $3/2$ times bigger than
$2 \s_3 / \s_2$, the correct coefficient of $f$ for $\var^2_r (f)$,
as shown in Eq.\eqref{E:var_ffcm} or Eq.\eqref{E:var_fixed_f}.

\section*{Acknowledgements}
  This work was supported in part by 
  the Clinical and Translational Science Award UL1 RR024139 
  from the National Center for Research Resources, 
  National Institutes of Health.


\begin{thebibliography}{7}
\providecommand{\natexlab}[1]{#1}
\providecommand{\url}[1]{{#1}}
\providecommand{\urlprefix}{URL }
\expandafter\ifx\csname urlstyle\endcsname\relax
  \providecommand{\doi}[1]{DOI~\discretionary{}{}{}#1}\else
  \providecommand{\doi}{DOI~\discretionary{}{}{}\begingroup
  \urlstyle{rm}\Url}\fi
\providecommand{\eprint}[2][]{\url{#2}}

\bibitem[{Balakrishnan and Koutras(2002)}]{Balakrishnan2002}
Balakrishnan N, Koutras MV (2002) Runs and Scans with Applications. John Wiley
  \& Sons, New York

\bibitem[{Flajolet and Sedgewick(2009)}]{Flajolet2009}
Flajolet P, Sedgewick R (2009) Analytic Combinatorics. Cambridge University
  Press, New York, NY, USA

\bibitem[{Kong(2006)}]{Kong2006}
Kong Y (2006) Distribution of runs and longest runs: A new generating function
  approach. Journal of the American Statistical Association 101:1253--1263

\bibitem[{Kong(2009{\natexlab{a}})}]{Kong2009}
Kong Y (2009{\natexlab{a}}) Statistical distributions of pyrosequencing. J
  Comput Biol 16:31--42, \doi{10.1089/cmb.2008.0106}

\bibitem[{Kong(2009{\natexlab{b}})}]{Kong2009b}
Kong Y (2009{\natexlab{b}}) Statistical distributions of sequencing by
  synthesis with probabilistic nucleotide incorporation. J Comput Biol
  16:817--827, \doi{10.1089/cmb.2008.0215}

\bibitem[{Kong(2012)}]{Kong2012}
Kong Y (2012) Length distribution of sequencing by synthesis: fixed flow cycle
  model. J Math Biol \doi{10.1007/s00285-012-0556-3}

\bibitem[{Kong(2013)}]{Kong2013}
Kong Y (2013) Distributions of runs revisited. Communications in Statistics -
  Theory and Methods To appear

\end{thebibliography}



\end{document}